\documentclass{iaus}
\usepackage{graphicx}

\newcommand{\Mo}{\mbox{M$_{\odot}$}}
\newcommand{\Moy}{\mbox{M$_{\odot}$~yr$^{-1}$}}

\title[Triggered Star and Galaxy Formation in Tidal Tails] %% give here short title %%
{Tidal Dwarf Galaxies as Laboratories of Star Formation and Cosmology}

\author[Duc, P.-A. et al.]   %% give here short author list %%
{Pierre-Alain Duc, Fr\'ed\'eric Bournaud \& M\'ed\'eric Boquien$^1$%
 \affiliation{$^1$ AIM - Unit\'e Mixte de Recherche CEA - CNRS - Universit\'e Paris VII - UMR n$^\circ$ 7158 Service d'Astrophysique, CEA--Saclay, 91191 Gif-sur-Yvette, France \break email: paduc@cea.fr\\[\affilskip]}}

\pubyear{2006}
\volume{237}  %% insert here IAU Symposium No.
\pagerange{1}
\date{?? and in revised form ??}
\jname{Triggered Star Formation in a Turbulent ISM}
\editors{B. G. Elmegreen \& J. Palous, eds.}
\begin{document}

\maketitle

\begin{abstract}
Star formation may take place in a variety of locations in interacting systems: in the dense core of mergers, in the shock regions at the interface of the colliding galaxies  and even within the  tidal debris expelled into the intergalactic medium. 
Along tidal tails, objects may be formed with masses ranging from those of super-star clusters to dwarf galaxies: the so-called Tidal Dwarf Galaxies (TDGs).  Based on a set of multi-wavelength observations and extensive numerical simulations, we show how TDGs  may simultaneously be used as laboratories to study  the process of star-formation (SFE, IMF) in a specific environment and as probes of various cosmological properties, such as the distribution of dark matter and satellites around galaxies.

%% add here a maximum of 10 keywords, to be taken form the file <Keywords.txt>
\end{abstract}

\firstsection % if your document starts with a section,
              % remove some space above using this command.
%\section{Introduction}

% NOTE use of \upi in above paragraph and subsequently throughout paper.
% The Greek constant character pi should be upright.

\section{The various observed  types of  star--forming tidal objects}\label{sec:sfr}
Star-formation in colliding galaxies has mostly been studied in their inner most regions. There the accumulation of gas, previously funneled by various dynamical processes, and its further collapse may lead to intense circum-nuclear starbursts. A mode of spatially extended, probably shock--induced,  star-formation has also been observed for a long time at the interface of the interacting galaxies, but was only recently modeled (\cite{Barnes04}). Further out the detection of HII regions along tidal tails proved that stars may form well outside the disks of the parent galaxies. The puzzling discovery of even more distant star-forming regions, located in the intergalactic medium, at more than 100 kpc from any massive galaxy, has raised the question of their origin. Given these large distances and related timescales, the young intergalactic stars were born in situ.
The relatively high metallicity measured in the associated ionized gas (typically half solar to solar) indicate that, without any doubt, the gas fueling these intergalactic SF episodes had previously been pre-enriched in the disk of parent galaxies and had later been expelled by a dynamical process: tidal forces or, for systems belonging to dense groups or clusters, by ram-pressure.

This paper focusses here on the formation of stars and stellar systems in the outermost regions of colliding galaxies.
%the other modes of star formation in interacting systems have indeed been addressed in other papers presented during the conference (for instance by Struck et al. and  Gao et al.). 
Before presenting how star--formation may take place far from the parent galaxies, it is worthwhile to briefly enumerate the various types of external star--forming objects so--far described in the literature:

\begin{itemize}
\item {\it Intergalactic emission--line regions}, with optical spectra  typical of star-forming HII regions.
They usually consist of small, often compact -- they are also refereed as  {\it EL-Dots} -- blue condensations made of a few  OB stars. Their  Star Formation Rates are below 0.01 \Moy\ (\cite{Gerhard02,Ryan-Weber04,Mendes04, Cortese06}).
 They will probably not form gravitationally bound objects but may contribute to the population of Intergalactic Stars, recently found in nearby clusters of galaxies.
 
 \item {\it Young (Super) Star clusters},  born in  giant HII complexes  and located along tidal tails  (e.g., \cite{Weilbacher03, deGrijs03,Lopez04}). Having typical typical masses of $10^{6-7} ~\Mo$,  the latter may evolve into {\it Globular Clusters}  (e.g. \cite{Schweizer96}).

\item {\it Tidal Dwarf Galaxies}, i.e. star--forming, gravitationally bound, objects made of tidal material, with apparent  masses and sizes of dwarf galaxies. The most massive of them exceed  $10^9~\Mo$ and are usually found close to the end of tidal tails. They contain large quantities of gas in atomic, molecular and ionized form (\cite{Braine01}).

\end{itemize}

\begin{figure}
 \includegraphics[width=\textwidth]{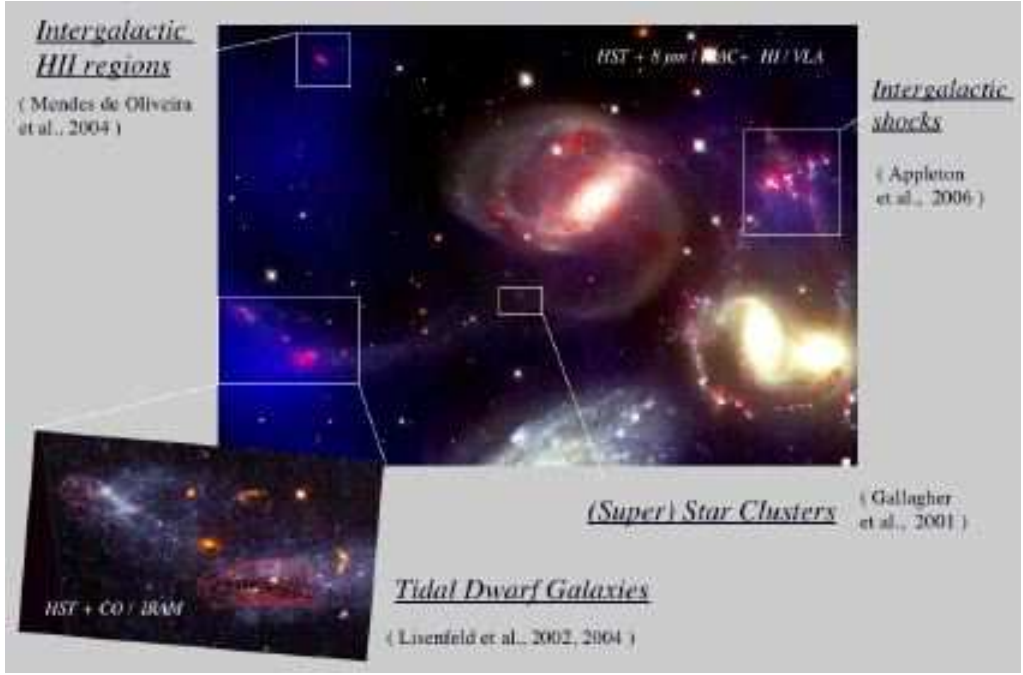}
  \caption{Various modes of Intergalactic Star Formation in the Stephan's Quintet. The main image is a montage of the Spitzer/IRAC 8 $\mu$m emission (in red), tracing the star--forming regions, the VLA HI emission (in blue), showing the gas reservoirs,  superimposed on an optical HST image, showing the stellar populations and dust lanes}\label{fig:sq}
\end{figure}

The variety of external star--forming regions/objects is perfectly illustrated  by the Stephan's Quintet (HCG92), a well studied compact group of galaxies (see Figure~\ref{fig:sq}). Instances of intergalactic HII regions (\cite{Mendes04}), young Star Clusters along tidal tails (\cite{Gallagher01}), Tidal Dwarf Galaxies (\cite{Lisenfeld04}) and even, probably shock induced Star Formation (\cite{Appleton06}), were already reported in the intragroup medium of HCG92.

\section{The formation and evolution of sub-structures along tidal tails, according to simulations}\label{sec:struct}

Depending on the characteristics of the star--forming regions, various stellar objects will or will not be formed in tidal debris. Numerical simulations may help in understanding the origin and the evolution of sub-structures along tidal tails. 

Soon after the study by \cite{Mirabel92} of a TDG candidate in the Antennae system, \cite{Barnes92} 
and  \cite{Elmegreen93}  published  numerical models exhibiting bound condensations 
along tidal tails with typical masses of $10^{6-8}$~\Mo. While in the  \cite{Barnes92} simulations, the condensations formed
from gravitational instabilities in the stellar component, those of  \cite{Elmegreen93}
were produced from gas clouds, the velocity dispersion of which had increased due to the collision. 
The direct formation of bound stellar objects  in models with limited resolution  has been questioned by \cite{Wetzstein06}. On the other hand, they claimed that the gaseous condensations themselves were not numerical artifacts.  

For practical reasons, most  simulations of galaxy--galaxy collisions assume that the dark  matter halo around  the parent galaxies is truncated, since, at first approximation, the latter  does not play a major role in the shaping of tidal tails. However,  \cite{Bournaud03} found that the internal structure of tidal tails does actually depend on the size of the  DM halo. When the latter exceeds ten times the size of the optical disk, prominent concentrations of mostly gaseous matter,  with  masses exceeding   $10^{9}$~\Mo, may form near the end of the tails.
% while smaller ones are still present along the tails (but are not as numerous as in the case of the truncated DM haloes). 
Such accumulations of gas  resemble  those present in the HI maps of several interacting systems. \cite{Duc04}  further explained that the shape of tidal forces, within the potential well of an extended DM halo, were such that gas clouds, originally located in the outskirts of the parent galaxies,  can be pulled out far away from them without being diluted. In the truncated DM halo case, matter is stretched along the tails, preventing the formation of massive condensations.  One should note that the presence of extended DM haloes is a prediction of cosmological simulations which has had sofar only a few rather indirect observational confirmations.  However, 
having  an extended DM halo is a necessary but not sufficient condition to form massive tidal objects.
As shown by \cite{Bournaud06},  favorable geometrical parameters for the collision and the initial presence of extended gaseous disks  are also key ingredients. The mass ratio between the parent galaxies, which should not be too different, does also matter. 

\begin{figure}
 \includegraphics[width=\textwidth]{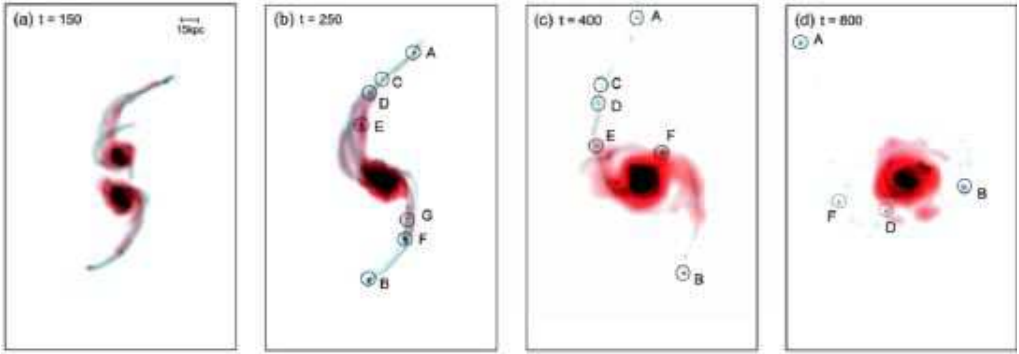}
  \caption{Numerical simulation of colliding galaxies. On each snap shot, the surviving tidal objects are identified (adapted from Bournaud \& Duc, 2006).}\label{fig:sim}
\end{figure}

Therefore, depending on their initial conditions, simulations predict the formation along tidal tails of a variety of objects with different origins: local instabilities along tidal tails, forming low or intermediate mass bound objects; a kinematic origin for the most massive ones. 
%Such a variety is also observed in real interacting systems. A commun characteristic is however that they form preferentially in the gaseous component, and are thus mostly made of young stars formed in situ. 
How these different types of objects evolve, how long they survive as independent bodies, are key questions that have sofar not had any observational answer, but may again be addressed with numerical simulations. 

Based on the analysis of  about 100 N--body simulations and  600 tidal objects (see one example on Figure~\ref{fig:sim}), \cite{Bournaud06}  found that, while most of the intermediate mass tidal objects along the tails vanished in a few 10$^8$ yrs, those initially located near the end of the tails, with typical masses of $10^9$ yrs, could survive up to at least 2 Gyr. They orbit around their parent galaxies like  satellite galaxies. The projected distribution of the long--lived tidal objects turns out to be remarkably similar to that of  the SDSS satellites around their host (\cite{Yang06}). In particular they show a similar anisotropy; the latter has probably a cosmological origin, but at least part of it could be due to a contamination by a population of Tidal Dwarf Galaxies.  The simulations by \cite{Bournaud06} did not have enough resolution to study the fate of the lowest--mass tidal objects. According to the ad--hoc simulations by \cite{Kroupa97}, those suffer severe gravitational evaporation,  loose a significant fraction of their mass, but some of them could still survive as low--mass dwarf spheroidals.  Other survivors may turn out to be the progenitors of the Super Star Clusters or even the Young Globular Clusters which were discovered  in interacting systems.

\section{Physical properties of  intergalactic star--forming regions}\label{sec:phys}

As shown before, Star Formation in interacting systems  may occur in a large variety of environments from the
densest ones to the most diffuse ones. It could  be expected that the modes of SF and thus the  Initial Mass Function (IMF) or the Star Formation Efficiency (SFE) would change from one type of regions to the other. Surprisingly,  \cite{Gao04} recently claimed that the SFE may not be so different in the dense core of Ultraluminous Infrared Galaxies and in the more quiescent regions  of spiral arms. This result urges the  comparison with the other extreme: the outermost regions of spirals and intergalactic star--forming regions. 

We have studied the Star Formation processes in the latter environment combining three popular indicators: the ultraviolet emission obtained with  Galex, which is sensitive to the SF episodes of time scales of about 100 Myr,  the H$\alpha$ emission, probing SF over time scales of less than 10 Myr and finally the mid--infrared emission obtained with Spitzer, which has a slightly higher timescale. 
%The far--infrared regime would provide a more direct SF indicator, but data in this wavelength domain will be rare until the launch of Hershel. All these indicators were calibrated and provide estimates of the Star Formation Rates.
Putting together the values of the Star Formation Rates estimated with each of these tracers, and comparing them with the amount of gas reservoirs,  one may determine the  dust obscuration and hence the total SFR and SFE as well as speculate on the starburst ages.  The  IMF and recent star formation history can further be reconstructed from the  analysis of the full Spectral Energy Diagram.
%Combining these multiwavelength data and constructing Spectral Energy Diagrams and comparing it with evolutionary synthesis models, one may even estimates the IMF.

\begin{figure}
 \includegraphics[width=\textwidth]{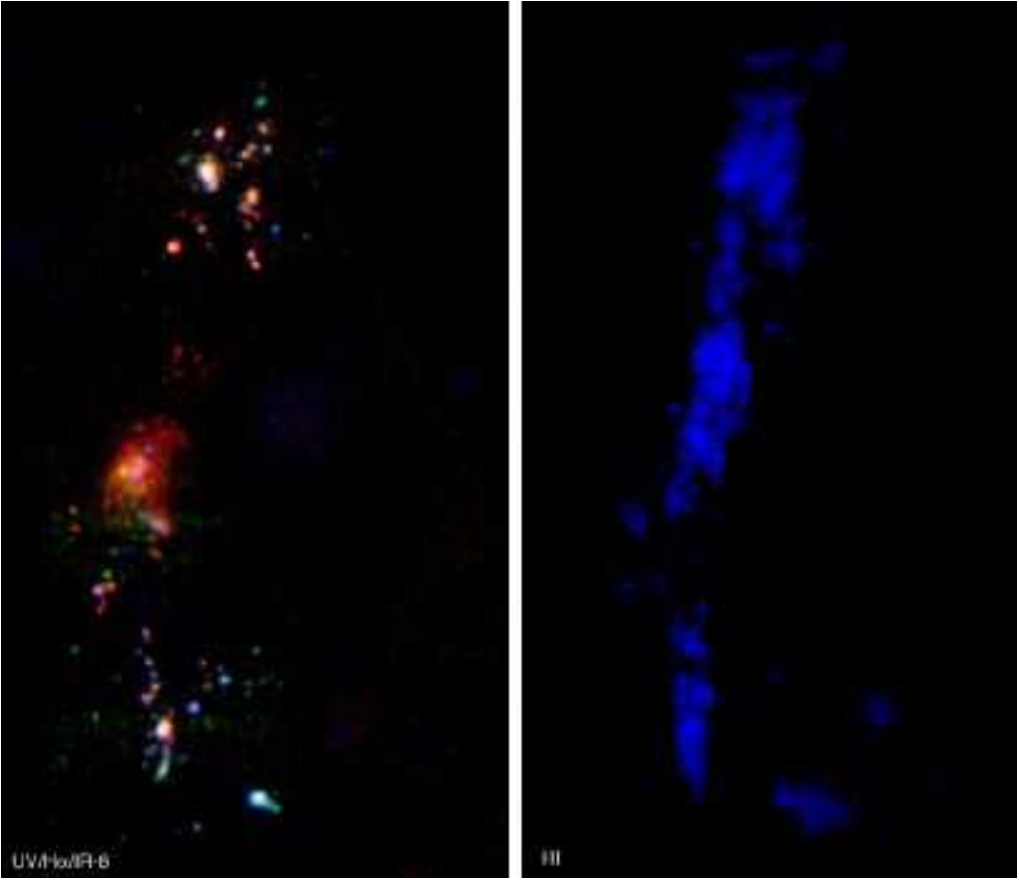}
  \caption{{\it Left}: composite image of the NGC 5291 system in pseudo-colours: near--ultraviolet Galex (blue), H$\alpha$ (green) and Spitzer/IRAC 8 $\mu$m (red), i.e. a combination of three star--formation indicators. {\it Right}: VLA-B array HI map with the same scale. The 30 intergalactic star--forming regions along the HI--ring like structure are characterized by a UV excess, best explained by fading, quasi--instantaneous starbursts ignited less than 10 Myr ago (Boquien et al., in prep.).  No evidence for an old stellar population has yet been found. }\label{fig:n5291}
\end{figure}

%We applied such a method to a set  of a few  interacting systems of particular interest for which we obtained either PI or archival Galex UV , Spitzer mid--IR  or  ground based H$\alpha$ data. 
The first  systematic studies of colliding galaxies with Spitzer indicate that globally the level of external star--formation, in particular  along tidal tails,  is rather low, contributing for less than 10\% to the total SFR (Struck et al. in this symposium). The objects we have selected, on the contrary, exhibit in their surroundings collisional debris that are particularly active.
For instance, in the interacting system NGC 5291 (see figure~\ref{fig:n5291} and Boquien et al., in these proceedings), more than 80\% of the current Star Formation occurs outside the main galaxies,  in about 30 individual intergalactic HII regions aligned along a huge 150--kpc long HI ring--like structure. The two most luminous ones have properties similar to TDGs, as defined in the previous section.

\section{Conclusions: TDGs as  laboratories of Star Formation and Cosmology}\label{sec:concl}

%\subsubsection*{Sar Formation}
Intergalactic SF regions, in general, Tidal Dwarf Galaxies, in particular,  appear as attractive laboratories to study the processes of star formation. Indeed, on one hand, they share with galactic SF regions the chemical characteristics -- at first order, they have the same ISM, including the molecular gas found in quantity in tidal tails --; on the other hand, they are by nature detached and isolated and are thus much simple to study.  Indeed, many of the observed intergalactic star forming regions were formed within "pure''  expelled gas clouds, with no evidence for the presence of a pre-existing stellar component from the parent galaxies. Without such a contamination,  the star--formation history (SFH) can be reconstructed with simple Single Star Population models.  In galactic disks where several generations of stars coexist, deriving the SFH is much more complex and ambiguous. Beside, for those SF episodes triggered by a tidal interaction, numerical simulations may provide dynamical ages for the system and thus upper limits on the SF onset time, provided it occurred after the formation and development of tidal tails. \\

%\subsubsection*{Cosmology}
Moreover, Tidal Dwarf Galaxies may be used to constrain some parameters related to CDM cosmology, even-though these second--generation galaxies should not contribute much more than 10\% to the overall dwarf galaxy population. First of all, TDGs with masses exceeding $10^{9}$ ~\Mo\, may only be formed if their parent galaxies were surrounded by an extended halo of dark matter, or more precisely if they were within a potential well causing flat rotation curves. Cosmological simulations predict that the halos made of non-baryonic  DM should be large; however their real sizes  are difficult to  measure directly. Second, because TDGs are made of material initially located in the disks of their parent galaxies, they should contain little quantities of the halo--type non baryonic dark--matter.  Thus, measuring their dynamical mass, and comparing it with the luminous one, should reveal the presence or absence of a yet unknown, baryonic, component of dark matter located in spiral disks. This would be a direct test for the existence of large quantities of, for instance, very  cold otherwise undetectable molecular clouds. Obviously deriving dynamical masses in objects as complex and young as tidal tails is still a challenge. This is nevertheless feasible in the most nearby systems (Braine et al., 2006, Bournaud et al., 2006 in prep.).
Finally, numerical simulations predict that the long--lived TDGs have orbits resembling those of satellite galaxies around their host. The number and distribution of the latter are critical as they  strongly constrain the cosmological, hierarchical,  models. The origin for the apparent  differences between the observed and predicted numbers of satellites has been actively debated for years. The existence of an additive population of dwarfs of tidal origin may even increase the discrepancies. 

One should keep in mind that many of the above conclusions relied on numerical simulations that, even-though they were thoroughly checked (in particular using several codes and numerical resolutions), should be confirmed with observations. Sofar, only young, still forming,  Tidal Dwarf Galaxies have been found and studied. They were classified as such thanks to  the tidal tail linking them to their parent galaxies. Older detached TDGs are obviously  much more difficult to pinpoint; unambiguous examples of such long-lived objects are still to be found.  The absence of a dominating dark matter halo, an unusually high metallicity,  are two hints for a tidal origin, which may be checked in nearby objects.

%\begin{acknowledgments}
%\end{acknowledgments}

%\begin{discussion}

\end{document}